\documentclass[twocolumn,aps,prc,superscriptaddress,showpacs]{revtex4}
\usepackage{amsmath,bm}%
\usepackage{graphicx}%

\begin{document}

\draft
%\preprint{}
\title{The calculation of total reaction cross sections induced by 
intermediate energy $\alpha$-particles with
BUU Model}
\thanks{Supported by the Major State Basic Research
Programme in China under Grant No. G2000077404}
\author{C. Zhong}
\thanks{E-Mail address: ZhongC@sinr.ac.cn}
\affiliation{Shanghai Institute of Nuclear Research, Chinese Academy of Scienecs, P.O. Box 800-204,
Shanghai 201800, China}
\author{X. Z. Cai}
\affiliation{Shanghai Institute of Nuclear Research, Chinese Academy of Scienecs, P.O. Box 800-204,
Shanghai 201800, China}
\author{W. Q. Shen}
\affiliation{Shanghai Institute of Nuclear Research, Chinese Academy of Scienecs, P.O. Box 800-204,
Shanghai 201800, China}
\affiliation{   CCAST (World Laboratory), PO Box 8730, Beijing 100080, China}
  \affiliation{  Department of Physics, Ningbo University, Ningbo 315211, China}
\author{H. Y. Zhang} 
\author{Y. B. Wei}
\author{J. G. Chen}
\affiliation{Shanghai Institute of Nuclear Research, Chinese Academy of Scienecs, P.O. Box 800-204,
Shanghai 201800, China}
\author{Y. G. Ma}
\affiliation{Shanghai Institute of Nuclear Research, Chinese Academy of Scienecs, P.O. Box 800-204,
Shanghai 201800, China}
\affiliation{   CCAST (World Laboratory), PO Box 8730, Beijing 100080, China}
\author{W. Guo}
\author{D. Q. Fang}
\affiliation{Shanghai Institute of Nuclear Research, Chinese Academy of Scienecs, P.O. Box 800-204,
Shanghai 201800, China}

\date{\today}
\begin{abstract}
The Boltzmann-Uehling-Uhlenbeck (BUU) Model, which includes the
Fermi motion, the mean field, individual nucleon-nucleon (N-N)
interactions and the Pauli blocking effect etc., is used to
calculate the total reaction cross section $\sigma_R$ induced by
$\alpha$-particles on different targets in the incident energy
range from 17.4 to 48.1 MeV/u. The calculation result can
reproduce the experimental data well. The nucleus-nucleus
interaction radius parameter $r_0$ was extracted from experimental
$\sigma_R$. It is found that $r_0$ becomes constant with
increasing the mass number of target.
\end{abstract}

\maketitle

%\nopagebreak 
The total reaction cross section, $\sigma_{R}$, is
one of the most fundamental quantities characterizing the nuclei
and the nuclear reaction. It can be used to deduce the nuclear
size and it relates with nuclear equation of state (EOS) and
in-medium (N-N) cross section also. It has been extensively
studied both theoretically and
experimentally.
\cite{Hussein1,Tan1,Tan2,Shen89,Cai00,Cai01,Ma1,Ma2,Ber88,
Mit1,Sai1,Oza1,Oza2,Feng1,Feng2,Fang99,Fang01,Vis1,Meng02}
There are, however, very few experimental results of the total
reaction cross sections, $\sigma_{R}$, for light-ions like $p$,
$d$, $^3$He, $\alpha$ etc. Recently $\sigma_R$ for
$\alpha$-particles, $^3$He, $d$ and proton on $^{9}$Be, $^{12}$C,
$^{16}$O, $^{28}$Si, $^{40}$Ca, $^{58,60}$Ni,
$^{112,116,120,124}$Sn, and $^{208}$Pb targets have been measured
at energies around several tens MeV/nucleon. It provided an
exciting chance to understand the mechanism of nuclear reaction
and give a more reliable test of different models. The
measurements were performed with a well-collimated and momentum
analyzed beam from the Gustaf Werner cyclotron at The Svedberg
Laboratory. The beam energy spread was approximated 100 keV
(FWHM), and the intensity was typically $2\times10^4$ particles
per second. A detailed description of the apparatus and
experimental technique is given in
Refs.\cite{Car1,Ing1,Ing2,Ing3,Auce1,Auce2}.

The results induced by light-ions are compared with predictions
from microscopic Glauber multile-scattering theory which is based
on the individual nucleon-nucleon (N-N) collisions in the overlap
volume of the colliding
nuclei.\cite{Vis1,Ing1,Ing2,Ing3,Auce1,Auce2,Ray1,War1,Maj1,Rid1}
This Glauber model calculation is a useful tool to study
$\sigma_{R}$. It considers the Coulomb correction, uses Yukawa
interaction with finite range force and distinguishes neutron and
proton inside nuclei. Comparisons of $\sigma_{R}$ at relativistic
energy with that at intermediate energies for projectiles heavier
than $\alpha$-particle have been done by Ozawa \emph{et
al}.\cite{Oza1,Oza2} The result calculated by the Glauber model is
always underestimated $\sigma_{R}$ at intermediate energies. This
problem has been solved now partly by few body Glauber model. For
light-ions like $p$, $d$, $^3$He, $\alpha$ etc., Glauber model can
fit experimental results, but the fit quality do not so well. Some
nuclear transport theories such as the Boltzmann-Uehling-Uhlenbeck
(BUU) model and quantum molecular dynamics (QMD) have been applied
into the calculation of $\sigma_{R}$ to resolve this problem
also.\cite{Cai00,Cai01,Ma1,Ma2,Ber88} These models incorporate the
Fermi motion, the mean field, individual nucleon-nucleon(N-N)
interactions and the Pauli blocking effect etc. in calculation.
They should be more suitable for total reaction cross section
calculation in intermediate incident energy. Test particles method
and the grid method originated from the fluid mechanics had been
introduced to revolve the BUU equation by C.Y. Wong et al. More
details can be found in Ref.\cite{Wong1}. For medium projectiles,
BUU model have been applied to calculate the total cross section
successfully,\cite{Cai00} So it is interesting whether BUU
calculation can be used in the calculation of total reaction cross
section induced by light ions like $\alpha$-particle.

In this letter we used the BUU model to calculate the total
reaction cross sections, $\sigma_{R}$, measured for 17.4 to 48.1
MeV $\alpha$-particles on targets from $^{9}$Be to
$^{208}$Pb.\cite{Ing2,Auce1} Within the framework of BUU model and
according to Poisson Statistics, the average $N$-$N$ collision
number can be obtained as a function of the impact parameter $b$
by assuming a reasonable parameterization of $\sigma_{NN}$. The
probability of $n$ times $N$-$N$ collisions $T_n(b)$ the course of
nucleus-nucleus reaction can be easily obtained in BUU
calculation. The sum of $T_{n}(b)$ over $n(n \ge 1)$ represents
the total probability of $N$-$N$ collisions and is related closely
to the absorption probability of nuclear reaction. Therefore, the total
reaction cross section is given by
\begin{equation}
\sigma_{R}=2\pi\int\left[\sum_{n=1}^{\infty}T_{n}(b)\right]b\textrm{d}b
=2\pi\int\left[1-\textrm{exp}(N)\right]b\textrm{d}b
\label{EQU04}
\end{equation}
where $N$ is the average $N$-$N$ collision number. More details
can be found in Ref.\cite{Ma1}. The BUU equation can be used to
extract nuclear equation of state (EOS) and in-medium $N$-$N$
cross section ,$\sigma^{in-medium}_{NN}$, etc., from the analysis
of the experimental excitation function for the total reaction
cross section, $\sigma_{R}$, if we know the nucleon density
distribution for the projectile and target.\cite{Ma1,Ma2}
%\begin{figure*}[t]
%\includegraphics[width=18cm]{alpha+target.eps}
\begin{figure}
\includegraphics[scale=0.5]{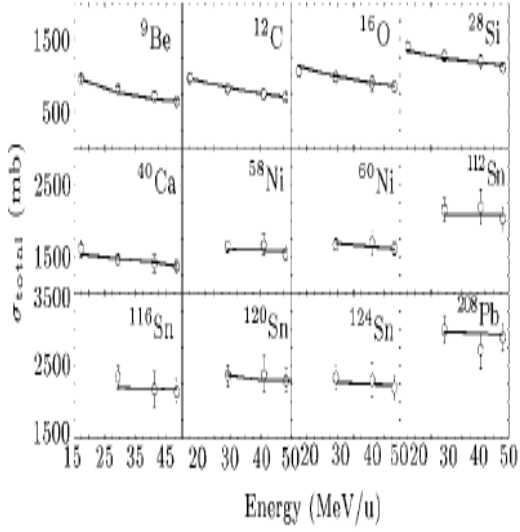}
\caption{Total reaction cross section induced by $\alpha$-particle
for various targets as function of incident energy. The experimental
data are indicted by open circles. Solid curves indicated the
calculations by using the BUU model.} \label{FIG1}
\end{figure}
The projectile and target nucleus are stable nucleus here in this
paper, so the well-known Cugnon's parameterization $\sigma_{Cug}$
for $\sigma_{NN}$\cite{Cug1}, soft EOS, and square-type density
distribution are used in the BUU calculation.\cite{Cai00} Then the
Nucleus-Nucleus interaction radius parameter $r_0$ can be deduced
by fitting experimental data at one incident energy, which is 48.1
MeV in this paper. As shown in Fig. 1, the experimental data are
indicated by closed circles. Solid curves indicated the calculated
results by using the BUU model. As seen, the energy dependence
changes with the mass number of targets. These experimental data
induced by $\alpha$-particles are in very well agreement with our
calculated results for all targets at all experimental incident
energy, where the experimental data at incident energy 48.1MeV
were used in fitting procedure to get $r_0$. For light targets the
total reaction cross section decreases with increasing energy, in
agreement with what is expected from the energy dependence of
nucleon-nucleon scattering cross section. For medium and heavy
targets, however, the total reaction cross section remains more or
less independent of energy and geometrical effects seem to
dominate. For medium and heavy-targets at low energies, maybe,
Coulomb repulsion suppresses the increase of the total reaction
cross section with decreasing the incident energy. The
experimental results produced by heavier projectiles indicates
also that for light target the total reaction cross sections are
much more sensitive to the nucleon-nucleon interaction. The total
reaction cross sections induced by deuterons and protons shown
same tendency either for the experimental results or for BUU
calculation.

The dependence of the total reaction cross section on the atomic
mass of target and projectile has been parameterized in many
forms. According to the strong absorption model, the total
reaction cross section $\sigma_R$ can be represented in terms of
the interaction radius $R$ and nucleus-nucleus interaction barrier
$B$, as
\begin{equation}
\sigma_{R}=10\pi R^{2}(1-B/E_{C.M.}) (mb)
\label{EQU05}
\end{equation}
where $R$ in fm and $E_{C.M.}$ in MeV. Different parameterized
formulas will have different parameterized forms for $R$ and $B$.
Shen \textit{{\it et\ al}} \cite{Shen89} propose a unified parametrized
formula for $\sigma_R$ using:
\begin{equation}
B=\frac{1.44Z_rZ_p}{r}-b\frac{R_tR_p}{R_t+R_p} (MeV)\label{EQU08}
\end{equation}

\begin{eqnarray}
R=r_0\left[A_t^{1/3}+A_p^{1/3}+1.85\frac{A_t^{1/3}A_p^{1/3}}{A_t^{1/3}+
A_p^{1/3}}-C(E)\right]+\nonumber\\
\alpha\frac{(A_t2Z_t)Z_p}{A_tA_p}+\beta 
E_{C.M.}^{-1/3}\frac{A_t^{1/3}A_p^{1/3}}{A_t^{1/3}+A_p^{1/3}}
\end{eqnarray}
where $r=R_t+R_p+3.2fm,$ $b=1Mev\cdot fm^{-1},$
$R_I=1.12A_i^{1/3}-0.94A_i^{1/3},(i=t,p).$ 
$\alpha=1fm,\beta=0.176MeV^{1/3}\cdot fm$
and the experience value around $1.0fm$ was used normaly for $r_0$.
%is calculated by Eq(2).
The nucleon radius parameter $r_0$ was also extracted from $\sigma_{R}$
calculated by BUU for 48.1MeV incident energy. As seen, this radius 
decreases with the mass number
for light nuclei, but it stays essentially constant at around 1.0 fm
for heavy nuclei, as shown in Fig. 2.
%\begin{figure}[t]
%\includegraphics[width=18cm]{r0-A_BUUcompShenFormula.eps}
%\caption{The nucleus-nucleus interaction radius parameter $r_0$ as
%function of target mass. The open stars denoted the calculations
%by using the BUU model. The open circles denoted the calculations
%by using the Eq(2-3-4).} \label{FIG2}
%\end{figure}
\begin{figure}
\includegraphics[scale=0.5]{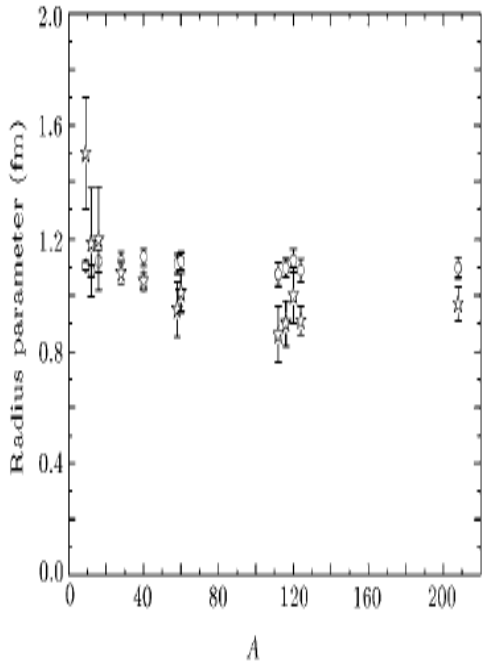}
%\begin{figure}[t]
%\includegraphics[width=18cm]{r0-A_BUUcompShenFormula.eps}
\caption{\footnotesize The nucleus-nucleus interaction radius parameter $r_0$ as
function of target mass. The open stars denoted the calculations
by using the BUU model. The open circles denoted the calculations
by using the Eq(2-3-4).}
\label{fig2}
\end{figure}
The open stars denoted the calculations by using the BUU model.
And the open circles denoted the calculations by using the
Eq(2-3-4) for 48.1MeV incident energy also. The deduced $r_0$ via
BUU can fit the results deduced from the parametrized formula of
Shen. It seems that the deduce $r_0$ shows similar trend with the
parametrized formula calculations. For $^{9}$Be, it has
two-$\alpha$ plus one neutron structure, so the radium is larger.
Our calculation shows this enhance of radium. S.Q. Zhang and J.
Meng propose a  unified parametrized formula for the nuclear
charge radii using Eq(5-6)\cite{SQZhang02}. Obvious Eq.6 is
exactly the isospin-dependent $A^{1/3}$ formula. It is interesting
to study the isospin-dependent of the target and the nuclear
charge radii in the future.
\begin{eqnarray}
R_{c} & = & r_{Z}Z^{1/3}=r_{Z}(\frac{A}{2}-\frac{N-Z}{2}) \nonumber \\
      & = & \frac{r_Z}{2^{1/3}}(1-\frac{N-Z}{A})^{1/3}\label{EQU09}
      \\
      & \approx & \frac{r_Z}{2^{1/3}}(1-\frac{1}{3}\frac{N-Z}{A})\label{EQU010}
\end{eqnarray}
%\begin{figure}[t]
%\includegraphics[width=18cm]{Be9-C12-Pb208_Test_Particles.eps}
\begin{figure}
\includegraphics[scale=0.5]{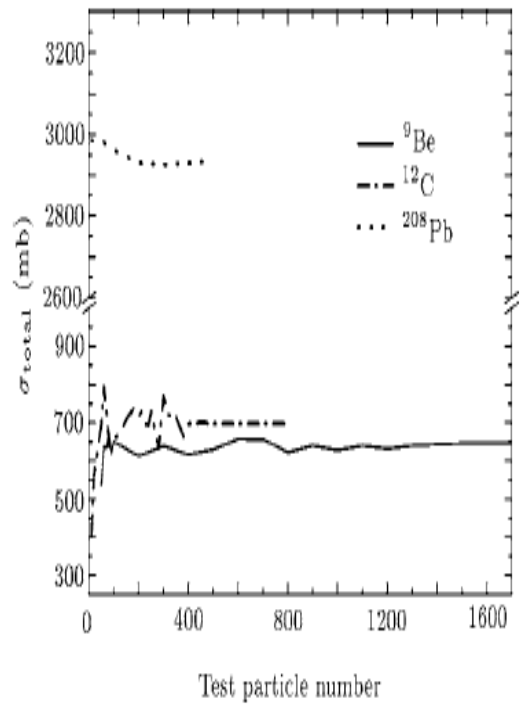}
\caption{Total reaction cross section induced by $\alpha$-particles as
function of the test particles. The solid line denoted $^9$Be target
The dash-dot denoted $^{12}$C target and the dot denoted $^{208}$Pb 
target.}
\label{FIG3}
\end{figure}

The total reaction cross section calculated by BUU is sensitive with the 
test
particles number. An important peculiar characteristic of the the test particles
simulation is to allow many test particles to represent one nucleon so as to gain a clear
insight into the dynamics of nuclear systems in heavy-ion dynamics.
In order to assess the test particles simulation as a steady and useful
concept, we wish to carry out some numerical calculations and compare them.
We examined the dynamics at 48.1MeV. The calculated results, here it is
the total reaction cross section, should reach stable saturation with
increasing the test particles number.
In our calculations, the calculation with different test particles was
carried out and the minimum test particle number is obtained. As shown
in Fig. 3, we use different numbers test particles to describe the
dynamics. For $^9$Be, in order to get the steady total reaction cross section
the test particles are 1000 at least. And for $^{12}$C, 400 test
particles are enough. For $^{208}$Pb, 200 test particles are enough even.
It shows that we can get steady $\sigma_{total}$ for $\alpha$-particle as
projectile and very light target as $^9$Be when the test particles numbers
is large enough.

In summary, we have presented the BUU Model calculation for the total
reaction cross sections induced by $\alpha$-particles at intermediate
energies on targets from $^{9}$Be to $^{208}$Pb. The BUU calculations
can fit the experimental data very well. The nucleus-nucleus interaction
radius parameter $r_0$ have been deduced from the BUU's calculation.
It is found that $r_0$ becomes constant with increasing the mass number
of target and shows similar trend with the parametrized
formula calculations. It shows that the total reaction cross sections
induced by light-ion like $\alpha$-particle can be calculated by the
BUU Model. Although the projectile has few nucleons, the BUU calculation
can be used still for the total reaction cross section calculation if large
enough test particle number was used.
Because test particle method, here in this paper different test 
particles
were used
The minimum test particle number ,which is needed for simulating one 
nucleon
in order to get back enough statistics and fit the experimental total 
reaction
cross section, is a function of the target mass number, it decrease with
increasing the target mass number.

This work is supported by the Major State Basic Research
Development Program in China under contract No. G200077400 and by
the National Natural Science Foundation of China under contract
10125521.

\end{document}